\documentclass[a4paper,longbibliography,twocolumn,superscriptaddress,showkeys,pre, aps, floatfix]{revtex4-1}

\usepackage[utf8]{inputenc}
\usepackage[english]{babel}
\usepackage{amssymb}
\usepackage{amsmath}
\usepackage{graphicx}
\usepackage[caption=false]{subfig}
\usepackage{float}
\usepackage[normalem]{ulem}
\usepackage{color}
\usepackage{placeins}

\usepackage[percent]{overpic}

\newcommand{\eqdot}{\,.}
\newcommand{\eqcomma}{\,,}

\renewcommand{\subsubsection}[1]{}

\newcommand{\piclab}[2]{
\begin{overpic}[width=0.32\textwidth]{#1}
 \put (10,78) {\large #2)}
\end{overpic}
}

\begin{document}

\author{Kaj-Kolja Kleineberg}
\email{kkleineberg@ethz.ch}
\affiliation{
Computational Social Science, ETH Zurich, Clausiusstrasse 50, CH-8092 Zurich, Switzerland}
\author{Dirk Helbing}
\affiliation{
Computational Social Science, ETH Zurich, Clausiusstrasse 50, CH-8092 Zurich, Switzerland}
\date{\today}

\title{Collective navigation of complex networks: Participatory greedy routing}

\begin{abstract}
Many networks are used to transfer information or goods, in other words, they are navigated. The larger the network, the more difficult 
it is to navigate efficiently.
Indeed, information routing in the Internet faces serious scalability problems due to its rapid growth, recently accelerated by the rise of the  Internet of Things. Large networks like the Internet can be navigated efficiently if nodes, or agents, actively forward information based on hidden maps underlying these systems. 
However,  
in reality most agents will deny to forward messages, which has a cost, and navigation is impossible.
Can we design appropriate incentives that lead to participation and global navigability?
Here, we 
present an evolutionary game where agents share the value generated by successful delivery of information or goods. 
We show that global navigability can emerge, 
but its complete breakdown is possible as well.
Furthermore, we show that the system tends to self-organize into local clusters of agents who participate in the navigation.
This organizational principle can be exploited to favor the emergence of global navigability in the system.
\end{abstract}


\maketitle

\section{Introduction}

The rapid growth of the Internet of Things is expected to lead to more than $50$ billion connected devices by $2020$~\cite{wehyper}, which implies challenges for the scalability of the digital infrastructure~\cite{internet:challenge,future:internet:broad,Godfrey:2009:PR:1594977.1592583,Gammon2010}.
Decentralized systems~\cite{helbing:digital_democracy,Contreras11122015,isocial} provide the required scalability in addition to the advantage of high transparency and a low risk of abuse by single, powerful entities~\cite{Epstein18082015,Bond2012}. 
These architectures, however, face particular problems concerning their functionality. 
A key challenge is how to route information relying only on local knowledge~\cite{Krioukov2010,Boguna2010,frag:hypermap,frag:hypermap_cn,Serrano2008,boguna:popularity,Gulys2015}.
To accomplish this task, the alignment of incentives of the involved individuals or institutions in the routing process to achieve global navigability is of crucial importance~\cite{internet:challenge}.

Routing information is among the most important functions real complex networks must perform. 
Examples include biological networks like the human brain, social networks~\cite{our:model}, and the Internet~\cite{Boguna2010,geometry:multilayer}. Many real complex networks have shown to be navigable, where nodes efficiently route messages using the connectivity of the network without relying on knowledge about the global topology. 
This can be achieved by performing \emph{greedy routing}~\cite{Papadopoulos2010,Boguna2008} in underlying geometric spaces~\cite{Krioukov2010,Boguna2010,frag:hypermap,frag:hypermap_cn,Serrano2008,boguna:popularity}. Greedy routing builds on nodes forwarding incoming messages to their neighbor that has the smallest distance to the destination measured in the underlying geometric space. 
 
In reality, there is a cost associated to sending messages, which can be a physical cost or invested time or energy. This cost implies that individuals are unlikely to be willing to participate without incentives. 
Furthermore, if even a single node in a given message forwarding chain decides not to participate (defect), the entire delivery fails. 
This vulnerability of the navigation process raises serious concerns about the feasibility of navigation in realistic environments. 
Here, we introduce an incentive to participate in the navigation process and show that
it can promote the emergence of global navigability. In reality, networks have a utility, and hence the successful delivery of information or goods generates value. We assume that this value is distributed equally among the individuals that took part in the forwarding process, which could be realized in digital environments by an appropriately designed cryptocurrency~\cite{socbit}. In the following we study an evolutionary game~\cite{maysmith,brownevo,PoncelaCasasnovas2016,Estrada2014,GmezGardees2012,Cardillo2014} that we call \textit{participatory greedy routing}. 
We show that in the real IPv6 Internet topology as well as in synthetic networks, the mentioned incentive allows for the emergence of global cooperation and navigability in the system despite its vulnerability to defection. On the other hand, complete defection and the total breakdown of navigability is also possible. 
We shed light onto the conditions under which the system approaches the navigable state. In addition, we reveal that global cooperation emerges as the system self-organizes into local clusters of cooperators, which tend to spread and merge. Finally, building on this knowledge, we show how the initial minimal density of cooperators needed to drive the system to the navigable state can be reduced significantly by adopting an ``act local'' strategy.

\begin{figure*}[t]
 \includegraphics[width=1\linewidth]{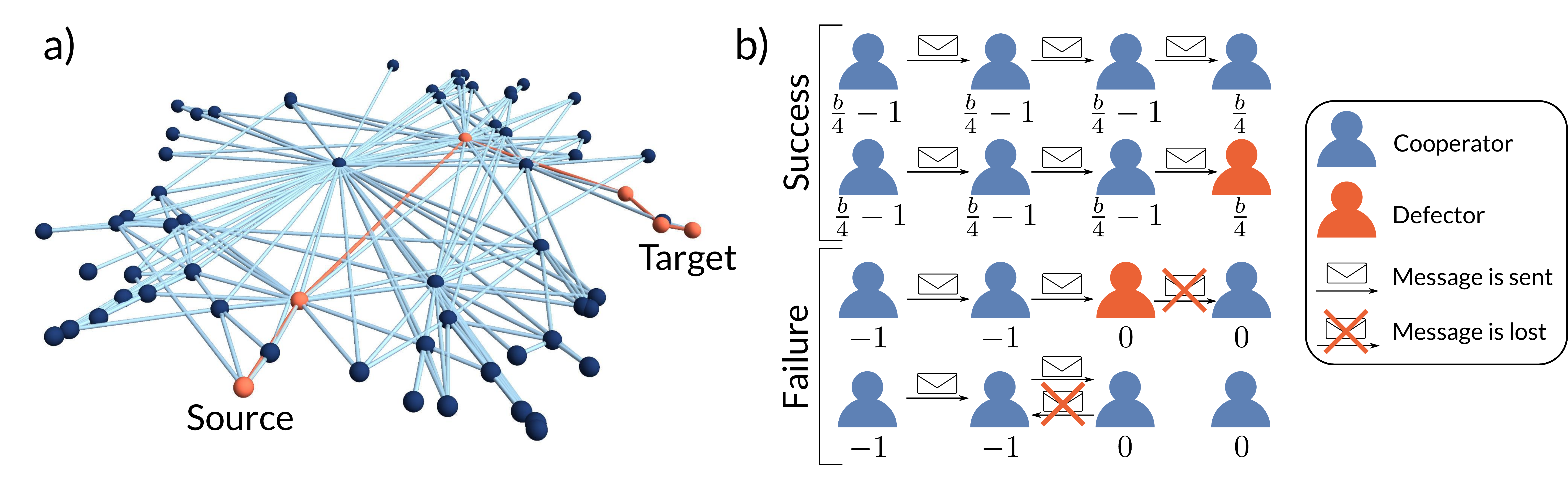}
 \caption{
 \textbf{(a)} Illustration of a message forwarding path found by greedy routing. Nodes and edges on the path are highlighted.
 \textbf{(b)} Illustration of the different possible scenarios of message forwarding events.
 In the first two cases, the message is delivered successfully. Each node obtains a share of the generated value $b$ and those that forwarded the message pay the cost of $1$. Here, we have four nodes involved (including the final recipient), and hence each node obtains the share $b/4$. 
 The state of the final destination has no impact, as a defector will also happily receive her message.
 In the third and fourth case the delivery fails. In the third case the message is given to a defector along the chain which simply does not forward the message. In the fourth case, although all nodes cooperate, the message runs into a loop and the delivery also fails, which is a property of the classical greedy routing procedure. In the third and fourth case, there is no reward, and each node that has sent the message to a neighbor pays the cost of $1$ (i.e. a payoff of $-1$). 
 \label{fig_sketch}}
\end{figure*}

\section{Results}

\subsection{Navigation of complex networks via hidden metric spaces}

It has been shown that real complex networks can be embedded into hidden hyperbolic metric spaces~\cite{frag:hypermap, frag:hypermap_cn}.
In these mappings, each node $i$ is mapped into the hyperbolic plane where it is represented by the polar coordinates $r_i, \theta_i$. These coordinates, or hidden variables, abstract the popularity and similarity of nodes~\cite{boguna:popularity}. The radial coordinate $r_i$ is related to the expected degree of node $i$ and therefore abstracts its popularity. More popular nodes are placed towards the center of the disc and hence obtain a lower radial coordinate. The angular distance between nodes $i$ and $j$ given by $\Delta \theta_{ij}=\pi-|\pi-|\theta_i-\theta_j||$ is a measure of the similarity of $i$ and $j$. A lower distance means a higher similarity. 
In simple terms, in a social network nodes close to you in this space would be your peers rather than your geographic neighbors. 
In the following, we always refer to this measure of proximity. 
The hyperbolic distance~\cite{Krioukov2010},
\begin{align} 
\begin{split}
x_{ij} & = \cosh^{-1}\left(\cosh{r_i}\cosh{r_j}-\sinh{r_i} \sinh {r_j} \cos{\Delta\theta_{ij}}\right) \\
 &  \approx r_i+r_j+2\ln{(\Delta\theta_{ij}/2)} \eqcomma
 \label{eqn_hypdist}
 \end{split}
\end{align}
comprises information about both the similarity and popularity of nodes $i$ and $j$. The connection probability for a given pair of nodes depends only on their hyperbolic distance. Given a real network, the coordinates of nodes can be inferred, using Maximum Likelihood Estimation techniques~\cite{Boguna2010,frag:hypermap, frag:hypermap_cn}.

In the inferred hyperbolic maps, one can identify soft communities, which are clusters of nodes close to each other in the angular similarity space~\cite{Boguna2010, Serrano2011, boguna:popularity, soft:comm}, predict missing links~\cite{Serrano2011,frag:hypermap, frag:hypermap_cn}, and perform efficient greedy routing reaching targets with high success rates and following approximatively topologically shortest paths~\cite{Boguna2010, frag:hypermap, frag:hypermap_cn, geometry:multilayer}. Importantly, the synthetic networks generated in this work are small-world networks~\cite{Watts1998}.

In greedy routing, each node forwards messages to her neighbor that is closest to the target in the underlying metric space. This means that node $i$ forwards the message to its neighbor node $j$ 
located at the smallest hyperbolic distance to the target node $t$, i.e. $x_{jt} = \min_{l \in \text{NB}(i)} x_{lt}$,
where $l$ runs over all neighbors of $i$ and $x$ is the hyperbolic distance from Eq.~\ref{eqn_hypdist}.
This process is illustrated in Fig.~\ref{fig_sketch}a.
Only local knowledge about the coordinates of neighbors and those of the target is needed to perform routing. If a message is given back to a node it has already visited, the delivery fails (see~\cite{Papadopoulos2010} for details). 

However, greedy routing assumes that all nodes, or agents, participate in the process, which is not the case in reality. In the following, we consider a cost associated with sending messages and present incentives that can lead to global-scale cooperation and a high performance of the system.

\subsection{Participatory greedy routing}

Participating in the forwarding of messages implies a commitment of the involved agents in terms of time and resources, in other words, sending messages has a cost. Individual agents will only be willing to pay this cost if they obtain a benefit from using the network. Hence, it is essential to create an incentive that can enable the emergence of global cooperation among the agents in the system. 
Real networks perform certain functions, and hence successfully delivering information or goods generates value. 
In decentralized systems, this value is generated in a bottom-up way by the agents who perform the routing. Consequently, we assume that the value generated, which we call $b$, is distributed equally among the agents that participate in the successful delivery.
In addition, each agent that sends a message has to pay a cost, which we set to $1$. This cost is paid even if the message is not delivered. Note that a single defector along the chain (except for the final destination) will lead to a failure of delivery (see Fig.~\ref{fig_sketch}b). Each forwarding chain is similar to a coordination game in the sense that the reward is only obtained if all nodes in the chain cooperate. However, 
costs are only paid until the message reaches the first defector. 

Is cooperation possible despite the strong impact that even a single defector can have? And if yes, under which conditions? To answer these questions, we design an evolutionary game called participatory greedy routing, in which individual agents choose their strategy such that they either cooperate or defect. Cooperators forward messages, whereas defectors ignore them and hence even a single one causes the entire delivery to fail. 

Participatory greedy routing is defined as follows and consists of three phases:
\textit{i) Initialization phase:} First, we either generate synthetic networks of size $N$ with an underlying hyperbolic geometry using the model described in Appendix~\ref{sec_app_h2}, or we use the real IPv6 Internet topology and its mapping to hyperbolic space (see Appendix~\ref{sec_appendix_ipv6}).
Then, we randomly distribute the initial cooperators, whose density we denote by $C_0$, in the network.
We will discuss the impact of a different distribution of initial cooperators later.
For now, let us move on to the
\textit{ii) navigation phase:} We randomly select pairs of nodes that act as source and target of a message sending process. We simulate the greedy routing process~\cite{Papadopoulos2010,Boguna2008} by forwarding the message to the neighbor closest to the target in the metric space at each step until it either reaches its destination or enters a loop or is passed on to a defector. In the last two cases, the delivery fails. After each message sending intent, we distribute the payoffs: each agents that has sent the message to another agent pays the cost of $1$ (i.e. her payoff is reduced by $1$), and only if the message was delivered successfully, each agent that participated in the forwarding chain (including the final destination) obtains her share $b / l_c$ of the generated value, where $l_c$ denotes the length of the chain including the final recipient.  
\textit{iii) Update phase:} After $N$ message sending intents, we perform 
a step where individuals can update their strategies following a replicator dynamics~\cite{evogamesgraphs,Cressman22072014,Helbing1996rep}.  
Each agent $i$ selects a random neighbor $j$ and copies her strategy (cooperate or defect) with a probability that depends on the payoff difference $p_j - p_i$, 
\begin{equation}
    p_{i \leftarrow j} = \frac{1}{1+e^{-(p_j-p_i)/K}} \eqcomma
\end{equation}
where $K$ mimics the randomness of decisions (here we always set $K=1$). After all agents update their strategy, we reset all payoffs and go back to the navigation phase until we reach the end of the simulation.

\begin{figure*}[t]
\piclab{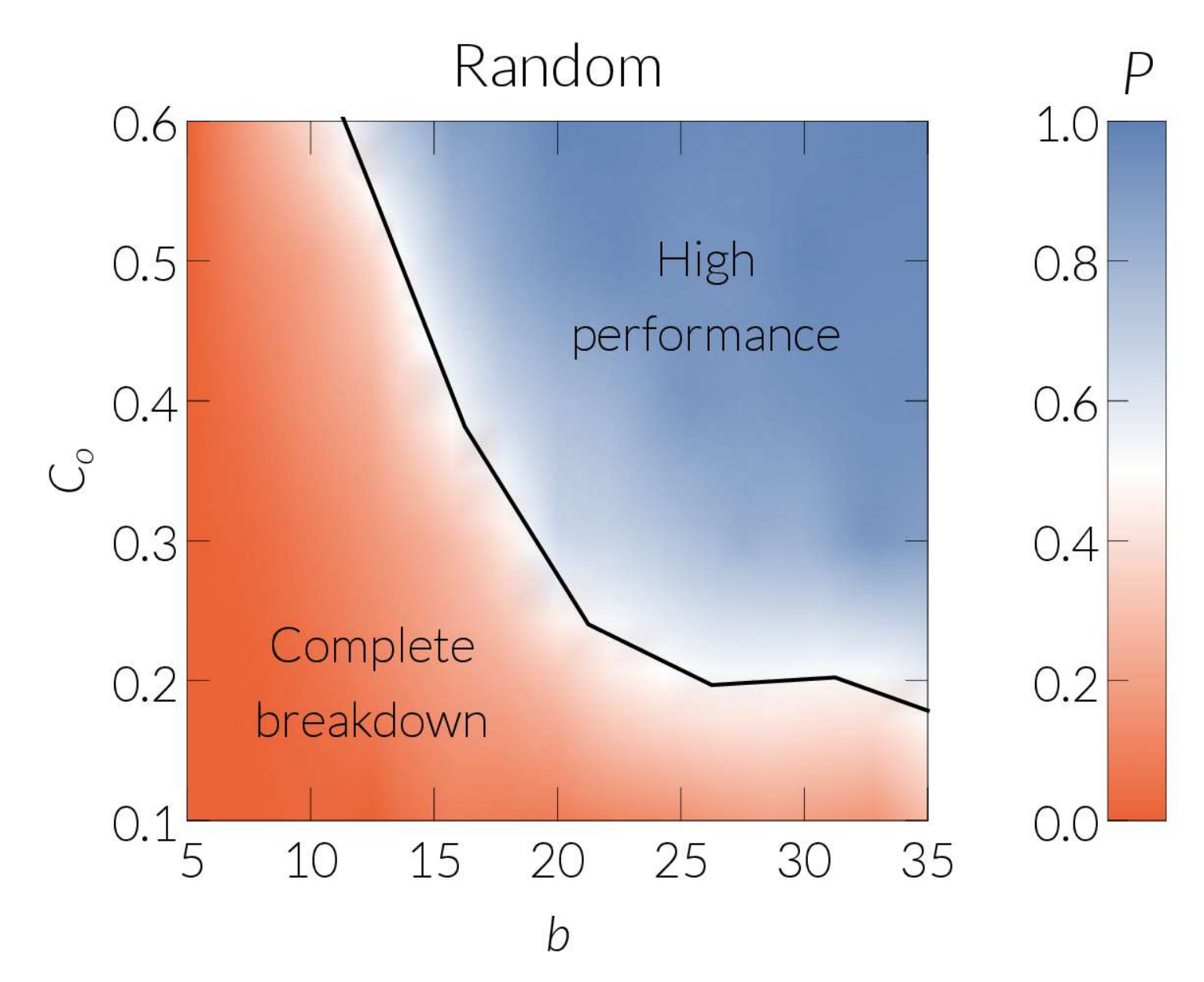}{a}
\piclab{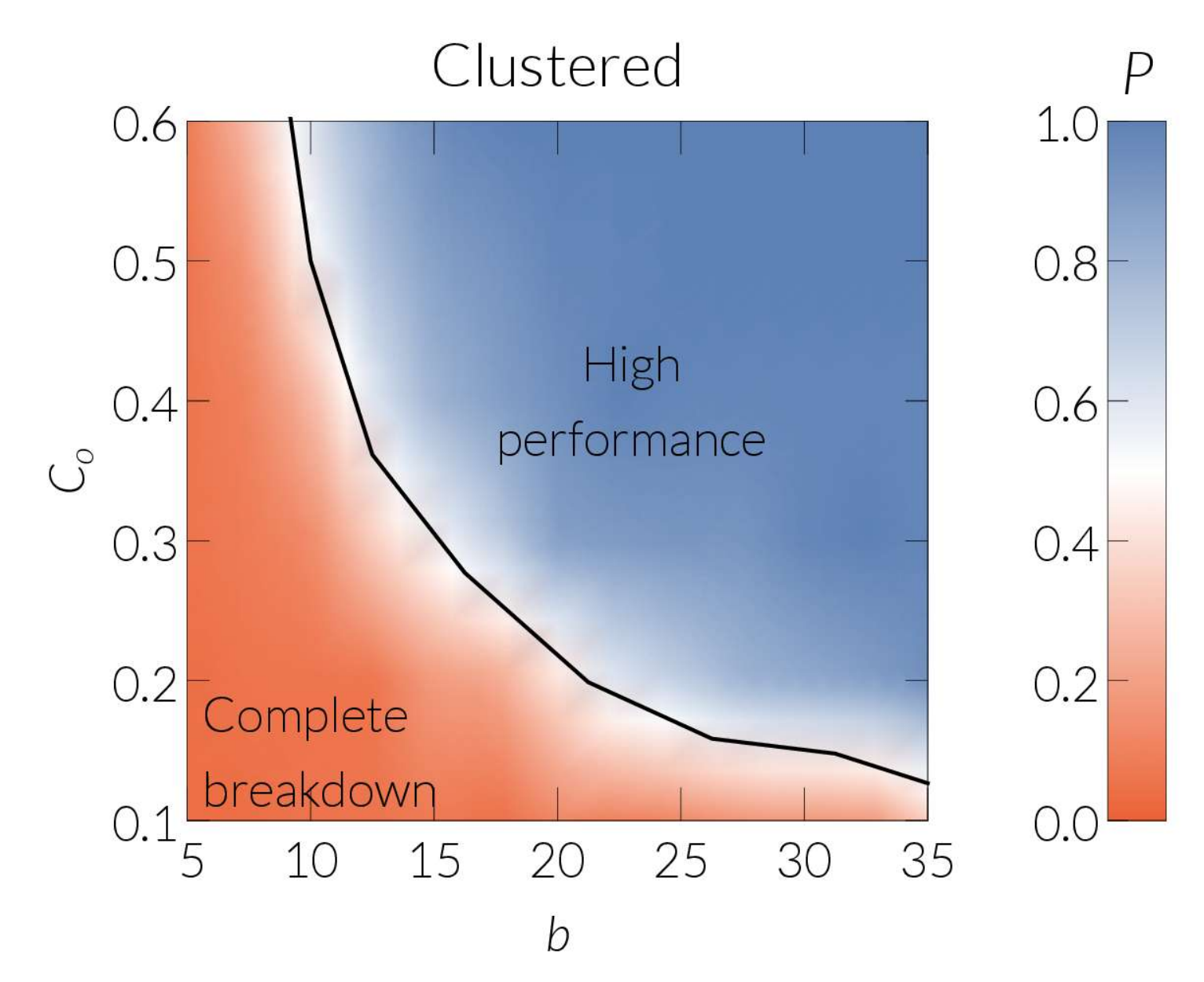}{b}
\piclab{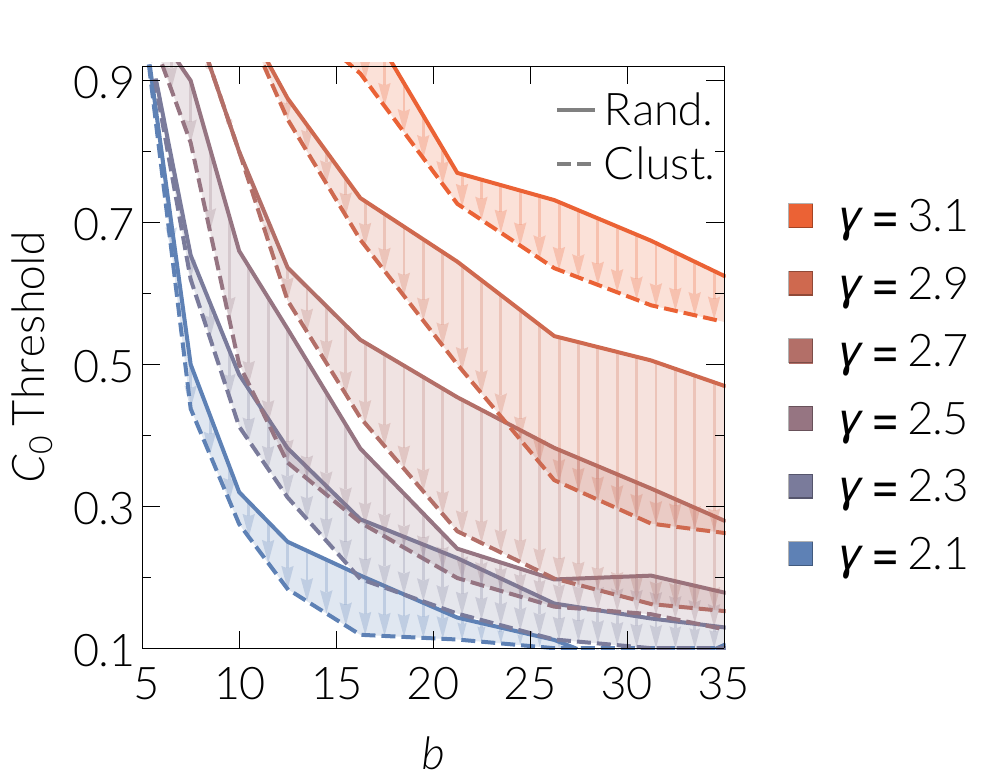}{c}
\\
\piclab{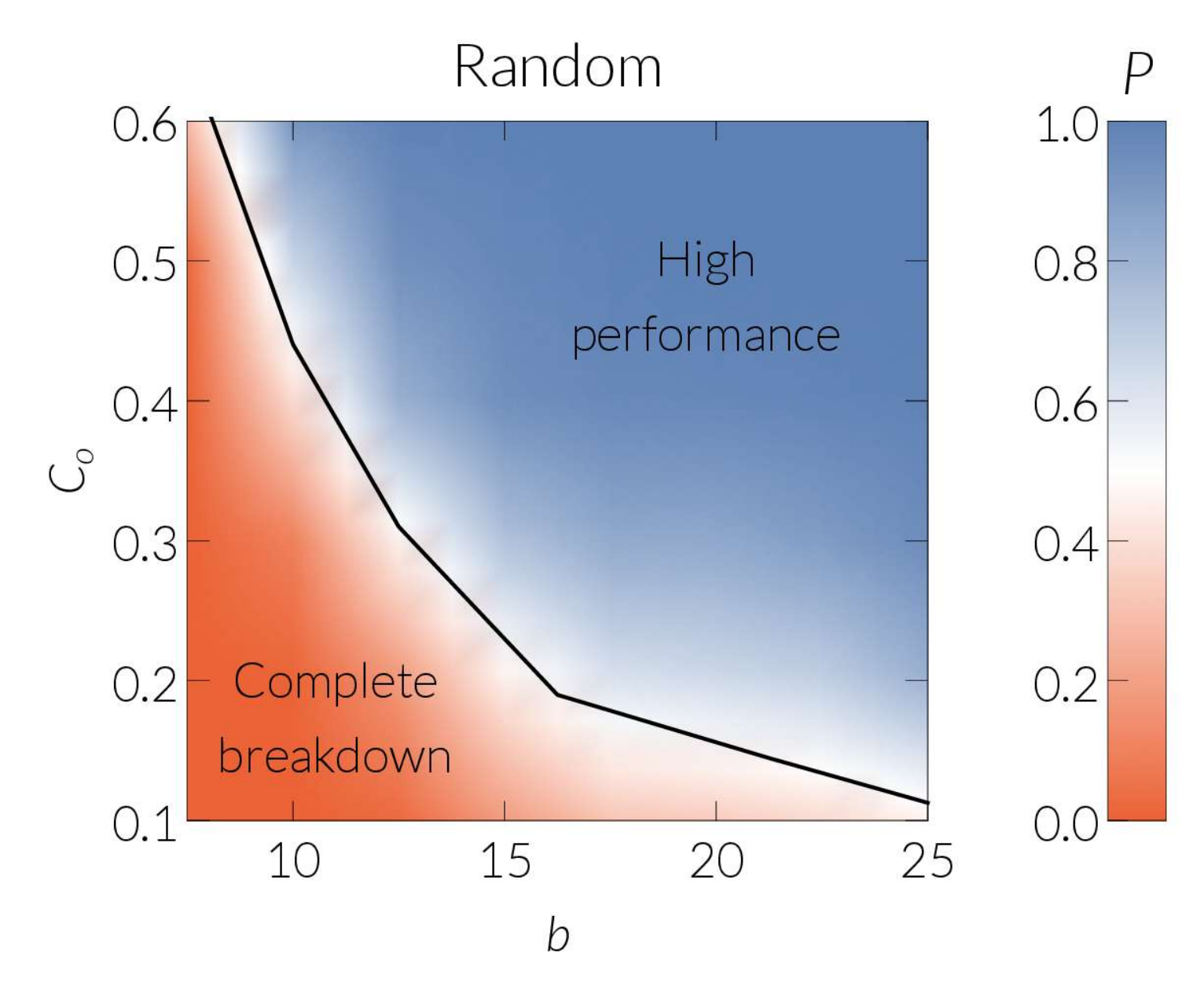}{d}
\piclab{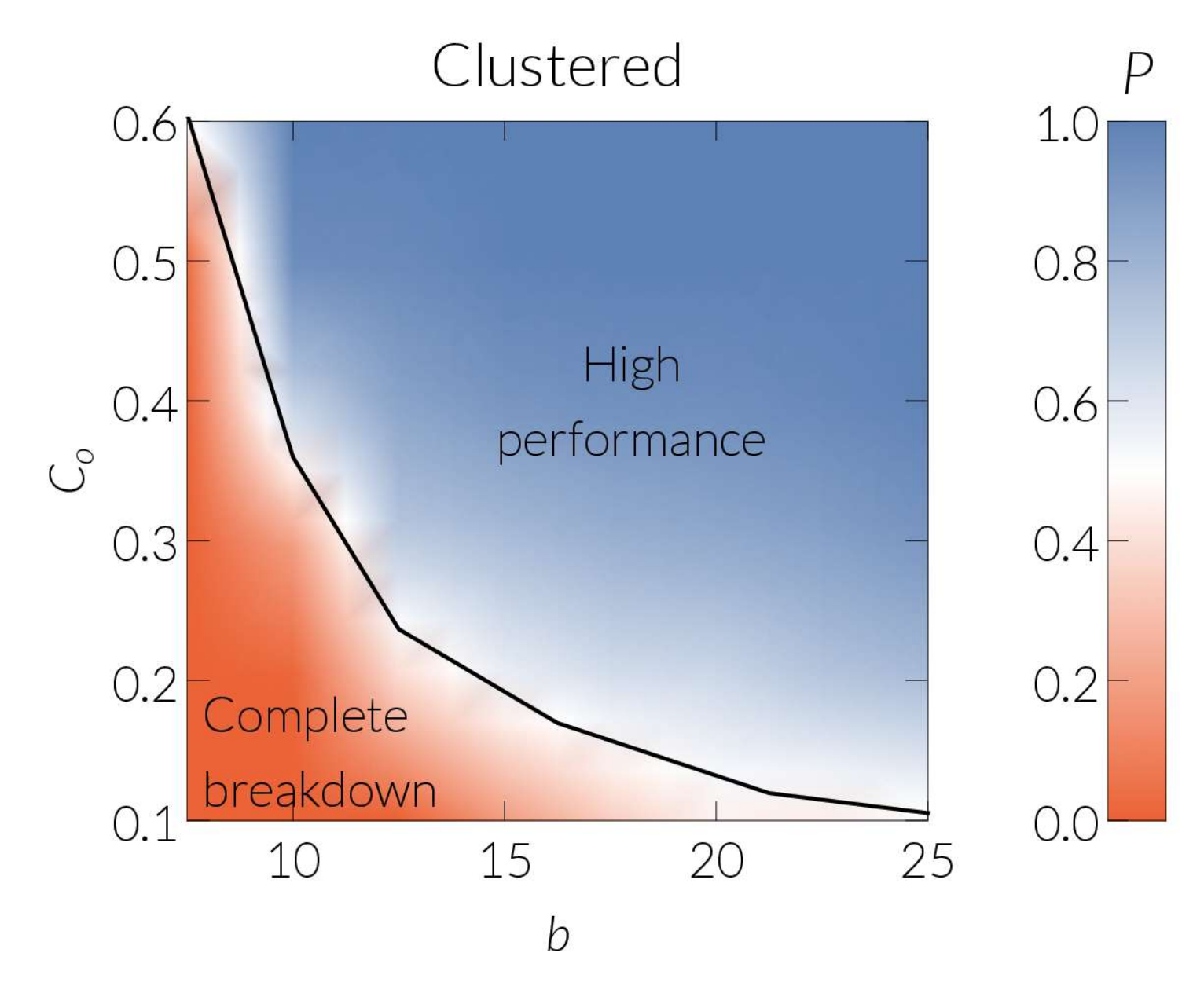}{e}
\begin{overpic}[width=0.24\textwidth]{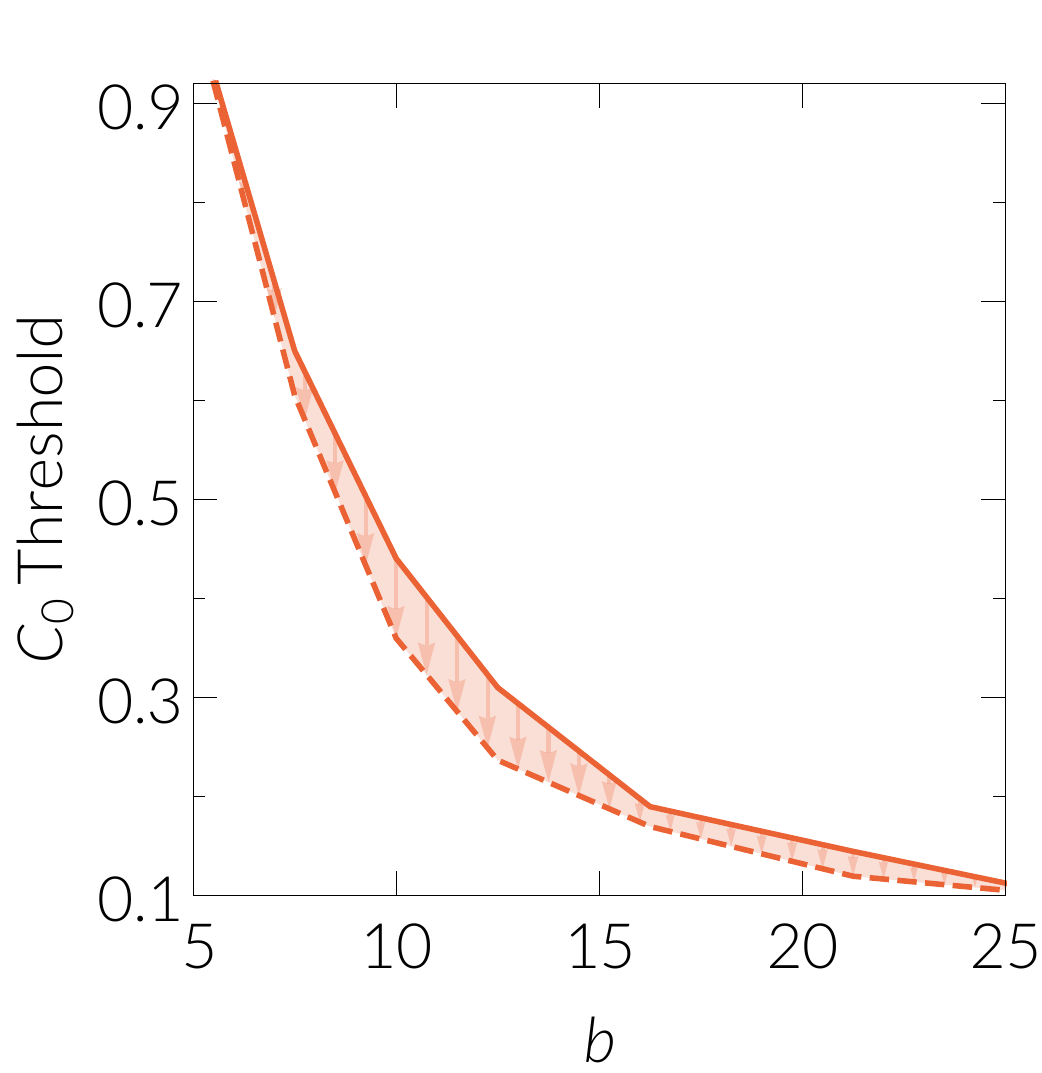}
 \put (10,100) {\large f)}
\end{overpic}
 \caption{
 \textbf{(a)} Probability of reaching the successful state (blue color) as a function of the initial cooperation rate $C_0$ and the payoff $b$. Here, we generated networks using the model described in Appendix~\ref{sec_app_h2} with $10000$ nodes, a power law degree distribution with exponent $\gamma = 2.5$, temperature $T=0.4$, and mean degree $\left< k \right> \approx 6$. 
 The results are averaged over $50$ simulations. A realization is said to have approached the functional state if the performance of the system in the stationary state is larger than $0.5$.
 To compute the quantities in the stationary state, we first let $250$ rounds pass, which avoids the initial transient phase, and then we average over another $250$ rounds. 
 Note that we only observe states that either perform very well with success rates $\gg 0.5$ or very poorly with success rates $\ll 0.5$ (see Supplementary Materials).
 Initial conditions are randomly assigned cooperators. \textbf{(b)} Same as before but starting from clustered cooperators according to Eq.~\eqref{eqn_localized}. 
 \textbf{(c)} Threshold value of $C_0$ for the system to approach the functional state with at least $50\%$ probability (black lines in \textbf{a, b}) for different values of the power-law exponent $\gamma$. Solid lines correspond to the random distribution of initial cooperators, whereas dashed lines represent the clustered initial configuration. Arrows show the reduction of the initial ``critical mass''. 
 \textbf{(d-e)} Same as \textbf{a-c} but for the IPv6 Internet network (averaged over $200$ realizations). 
 \label{fig_success_probability}
 }
\end{figure*}

\subsection{Bistability of the system}

We find that the system can approach two qualitatively different states in the long term. A functional state, in which nearly the entire population cooperates and the system is highly navigable, and a non-functional state, in which nearly the whole population defects and the system cannot be navigated at all (see Fig.~\ref{fig_success_probability}a). This bistability can be understood intuitively from the feedback between system performance and cooperation. If cooperation decreases, the system as a whole performs worse, and the temptation to defect increases, as obtaining the reward for successfully delivering a message becomes less likely. Hence, defection increases, which further decreases the system performance, which further increases defection, and so on, leading eventually to a complete breakdown of cooperation~\cite{helbing2013globally} and navigability. 
On the other hand, more cooperation means a better system performance, making cooperation a better strategy choice, which can lead to nearly full cooperation and navigability.  

The final fate of the system is determined by the initial rate of cooperators $C_0$ and the payoff parameter $b$.
We find that in a large region of the phasespace either the functional state (blue region in Fig.~\ref{fig_success_probability}a,b,d,e) or the non-functional state (red region) is approached with a high probability. These regions are separated by a smaller region in which both states are approached with intermediate probabilities (white region). 

Finally, note that, for full cooperation, participatory greedy routing degenerates towards greedy routing. 
In this optimal case, 
we find for the synthetic networks ($\gamma = 2.5$) a success rate of $(87.2 \pm 3.1) \%$ for greedy routing with an average of $(4.1 \pm 0.3)$ hops. For the Internet IPv6 network, we find that $(91.6 \pm 0.5)$ percent of the messages are delivered successfully with $(3.61 \pm 0.03)$ hops on average. In the functional state, participatory greedy routing yields a performance close to these optimal values (see Supplementary Materials).

\subsection{Self-organized local clusters drive global cooperation}

\begin{figure*}[p]
\includegraphics[width=0.9\linewidth]{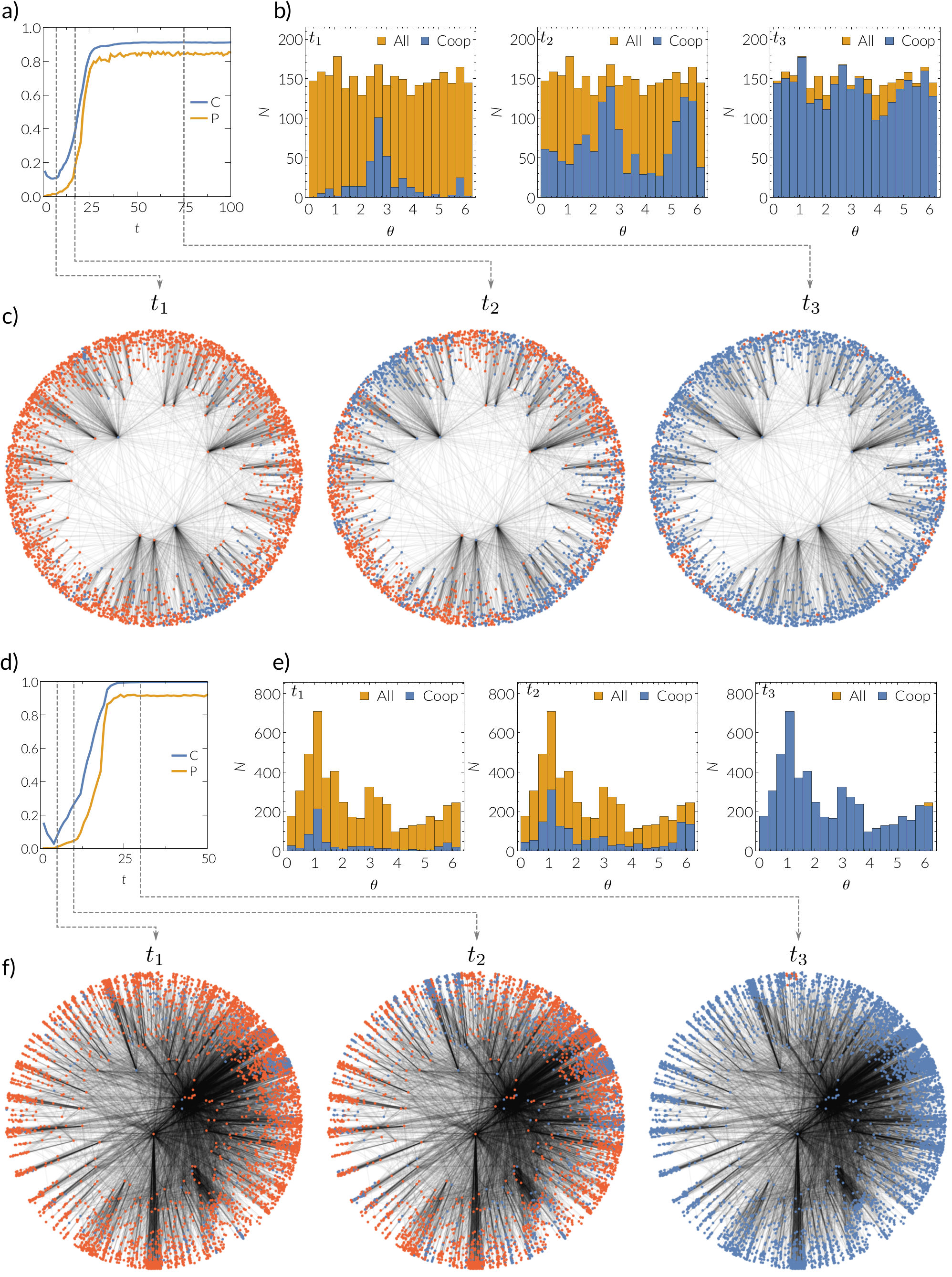}
 \caption{\textbf{(a)} Emergence of cooperation in a synthetic network for $b=25$ and $C_0 = 0.15$ with $N=3000$ nodes and a power law degree distribution with exponent $\gamma = 2.5$, temperature $T=0.4$, and mean degree $\left< k \right> \approx 6$.
 We show the evolution of the density of cooperators (blue line) and the success rate of navigation (yellow line). Time $t$ represents the number of update phases the system has undergone.
 \textbf{(b)} Histograms for the same realization for the number of cooperators (blue) and total number of nodes (yellow) in respective bins of the angular coordinate $\theta$. Time increases from left to right, $t= (7,17,75)$. \textbf{(c)} Network and states of the nodes for times $t= (7,17,75)$ for the same realization, see Supplementary Video 1. 
 \textbf{(e-f)} Same as \textbf{a-c}, but for the IPv6 Internet network and for $C_0 = 0.15$ and $b=30$. Times shown are $t= (5, 10, 30)$, see Supplementary Video 2. 
 \label{fig_panel}}
\end{figure*}

We observe that the spontaneous emergence of cooperating local clusters precedes the global emergence of cooperation~\cite{Helbing2009pnas}, and the underlying metric space offers a natural representation of such clusters. In Fig.~\ref{fig_panel} we present an explicit realization of the system for synthetic networks (see Supplementary Video 1) and one for the IPv6 Internet network (see Supplementary Video 2), where we start with a small fraction of cooperators that are chosen at random. Initially, the fraction of cooperators decreases (see Fig.~\ref{fig_panel}a), but at the same time localized clusters of high cooperation emerge. Such clusters are
sets of nodes in a small angular interval of the underlying space (see Fig.~\ref{fig_panel}b and c). 
Importantly, these clusters are successful because message paths follow shortest distances, and hence nodes within such a cluster can communicate despite the initially very low cooperation and poor performance of the system globally. If enough value $b$ is generated, these clusters can create a higher payoff than their defecting neighbors. This implies that defectors located at the border of a cooperating cluster have a high probability of switching to cooperation, which results in their assimilation of the cluster. Hence, the localized clusters of high cooperation grow and eventually merge until they cover a large fraction of the network. To sum up, in simple terms, global cooperation emerges from the spread of local clusters of cooperators.

\subsection{Heterogeneity and clustering of cooperators reduces the initial ``critical mass''}

We have seen that an initial ``critical mass'', i.e. a certain minimal density of initial cooperators is needed to drive the system to the functional state. We find that less initial cooperators are needed if the networks are more heterogeneous (see Fig.~\ref{fig_success_probability}c). The navigability of the system---even for full cooperation---decreases quickly 
for
$\gamma > 2.9$, which implies that it becomes very difficult to drive the system to the state of high cooperation (see Supplementary Materials). Note that most real complex networks have power-law exponents between $2$ and $3$ and are navigable~\cite{Boguna2010}.

As discussed in the previous section, the system dynamically self-organizes into localized clusters of high cooperation, which then expand and merge, eventually giving rise to global cooperation~\cite{Helbing2009pnas}. We can exploit the understanding of this process to lower the minimal initial density of cooperators. In particular, instead of distributing them randomly, we can assign them to a certain region of the angular space (see Supplementary Fig.~1f) by assigning to node $i$ the strategy 
 \begin{equation}
 S_i(0) = \begin{cases} 1 \quad \text{if} \quad \theta_i \in [0,2 \pi C_0), \\
        0 \quad \text{else},
       \end{cases}
       \label{eqn_localized}
 \end{equation}
 where $\theta_i$ denotes the angular coordinate of node $i$. This leads on average to an initial density $C_0$ of cooperators due to the uniform distribution of angular coordinates in the model for the generation of synthetic networks (see Appendix~\ref{sec_app_h2}). 
 This strategy significantly reduces the minimal initial density of cooperators required to reach the desired functional state. 
We define a threshold of initial cooperators above which the functional state is approached with more than $50\%$ probability. 
We find that starting with clustered initial cooperators can reduce this threshold by up to $50\%$ for the synthetic networks considered ($\gamma=2.7$), and by up to $24\%$ for the IPv6 Internet topology (see Fig.~\ref{fig_success_probability}c and f).
 To conclude, starting with local clusters of cooperators strongly favors the emergence of global cooperation and navigability.

\section{Community structure of the real Internet network}

As mentioned above, the real Internet IPv6 topology leads to similar results as for the case of synthetic networks.
Especially, we again find a bistable behavior (see Fig.~\ref{fig_success_probability}d and e), meaning that the system can be either in a state of high cooperation and navigability or nearly total defection and complete breakdown of navigability. 
We also find that clustering of the initial cooperators~\footnote{Note that
the non-uniform distribution of angular coordinates requires a slightly different way to assign localized initial conditions, which we explain in Appendix~\ref{app_sec_inet_initial}.} decreases the minimal density required to drive the system to the desired state (see Fig.~\ref{fig_success_probability}f). In accordance with our findings concerning synthetic networks, the higher heterogeneity present in the Internet IPv6 topology ($\gamma \approx 2.1$) favors the emergence of global cooperation and reduces the relative effect of clustered initial cooperators compared to distributing them randomly (compare Fig.~\ref{fig_success_probability}c and f). 

The existence of angular bins with significantly higher than average number of nodes is an indicator of a community structure that is present in real networks, in contrast to the synthetic networks considered here. In the real Internet IPv6 topology, the emerging clusters of high cooperation coincide---to some extend---with the community structure of the network (see Fig.~\ref{fig_panel}e, f, and Supplementary Video 2).

\section{Discussion}

Real complex networks have shown to be navigable efficiently and relying only on local knowledge by
performing greedy routing in underlying hidden metric spaces. However, greedy routing assumes that all nodes, or agents, participate in the process, which is often not the case in reality. Forwarding information has a cost for the agents, which can be a physical cost or invested time or energy. This cost is especially important because it causes a temptation for individual agents to defect (not participate in the navigation).
The navigation process is vulnerable to defection because even a single defector leads to the failure of an entire forwarding chain.
However, successfully delivering information in real networks generates value,
which in decentralized systems is generated in a bottom-up way. 

We assume that this value is shared among the agents that participate in the successful delivery. Agents can cooperate or defect, i.e. participate or not. We call this process participatory greedy routing.
We show in the real IPv6 Internet network as well as in synthetic networks that, in participatory greedy routing, global cooperation can emerge from local interactions. We show that the system is effectively bistable, and can either be functional, in which case nearly the entire population cooperates and the network is highly navigable, or non-functional, which corresponds to the complete breakdown of
navigability and nearly total defection. Interestingly, the emergence of global cooperation is preceded by the self-organization of the system into clusters of high cooperation. 
This organizational principle can be exploited to lower the critical initial density of cooperators needed to drive the system---with a high probability---to the desired functional state. 
We show that, if the initial cooperators are concentrated in a 
local cluster,  
the required minimal number of initial cooperators is reduced significantly.

The inclusion of a reputation system~\cite{Cuesta2015}, competition between several networks~\cite{geometry:multilayer,ecology20,worldmodel,layerlayer}, and adapting the network topology~\cite{Gulys2015} constitute interesting tasks for future work. 
Participatory navigation procedures like the one described in this manuscript are likely to play an important role in decentralized future digital environments and could sustain information routing in the Internet. 
Our results confirm that the emergence of global cooperation and navigability is possible with the correct incentives
and that the adoption of participatory bottom-up navigation is favored by starting with local clusters of participating agents. This is particularly important for the implementation in real systems.

\begin{acknowledgments}
We thank Bary Pradelski for interesting discussions about game theory and networks embedded in metric spaces.
D. H. is grateful for partial support by the ERC Grant “Momentum” (324247).
\end{acknowledgments}

\appendix

\section{Synthetic networks}
\label{sec_app_h2}

We construct synthetic networks using the $\mathbb{S}^1$ model described in~\cite{Serrano2008}, which we then map to its approximatively equivalent counterpart in the hyperbolic plane, the $\mathbb{H}^{2}$~\cite{Krioukov2010} model.

\subsection{$\mathbb{S}^1$ model}
In the $\mathbb{S}^1$ model each node $i$ has a set of hidden variables $\kappa_i, \theta_i$. $\kappa_i$ denotes the expected degree of node $i$, and the angular coordinate $\theta_i \in [0, 2 \pi)$ abstracts a similarity space. The model parameters are size $N$, the exponent for the power law degree distribution, $\gamma > 2$, average degree $\bar{k}$, and temperature $T \in [0,1]$. The network generation procedure then works as follows:
\begin{enumerate}
 \item Assign hidden variables: For each of the $N$ nodes, we draw $\theta_i$ from the uniform distribution in $[0, 2 \pi)$ and $\kappa_i$ from the probability density function 
 \begin{eqnarray}
\rho(\kappa) &=& (\gamma-1) \kappa^{\text{min}^{\gamma-1}} \kappa^{-\gamma},\\
\nonumber \kappa^{\text{min}}&=&\bar{k}\frac{\gamma-2}{\gamma-1},
\end{eqnarray}
where $\kappa^{\text{min}}$ controls the expected mean degree $\bar{k}$. 
\item Now, we connect each pair of nodes $i,j$ with the probability
\begin{eqnarray}
\label{r_s1}
r(\kappa_{i}, \theta_{i}; \kappa_{j}, \theta_{j})  &=& {1
\over 1 + \left[\frac{d\left(\theta_i ,\theta_j\right)}{\mu\kappa_i\kappa_j}\right]^{\frac{1}{T}}},\\
\nonumber d(\theta_i, \theta_j) &=& \frac{N}{2\pi} \Delta \theta_{ij},~\Delta \theta_{ij}=|\pi - | \pi -|\theta_i - \theta_j|||,\\
\nonumber \mu&=&\frac{\sin{T \pi}}{2\bar{k}T\pi},
\end{eqnarray}
where $d(\theta_i, \theta_j)$ denotes the the angular distance between $i$ and $j$.
\end{enumerate}

\subsection{Transformation to $\mathbb{H}^2$}
We transform the $\mathbb{S}^1$ model to the $\mathbb{H}^2$ model in the hyperbolic plane by mapping the hidden variables $\kappa_i$ to radial coordinates $r_i$ according to
\begin{equation}
\label{kappa_i}
r_i = R - 2 \ln \frac{\kappa_i}{\kappa^{\text{min}}},
\end{equation} 
where $R$ denotes the disc radius given by
\begin{eqnarray}
\label{R}
R&=&2\ln{\frac{N}{c}},\\
\nonumber c&=&\bar{k}\frac{\sin{T\pi}}{2T}\left(\frac{\gamma-2}{\gamma-1}\right)^2 \eqdot
\end{eqnarray}
Finally, note that in the $\mathbb{H}^2$ model the connection probability from Eq.~(\ref{r_s1}) translates to the 
Fermi-Dirac distribution 
\begin{equation}
\label{fermi_dirac}
p(x_{ij})=\frac{1}{1+e^{\frac{1}{2T}(x_{ij}-R)}}, 
\end{equation}
and depends only on the hyperbolic distance
\begin{equation} 
 x_{ij} = \mathrm{cosh}^{-1}\left(\cosh{r_i}\cosh{r_j}-\sinh{r_i} \sinh {r_j} \cos{\Delta\theta_{ij}}\right)
\end{equation}
between the nodes~\cite{Krioukov2010}.

\section{The IPv6 Internet topology}
\label{sec_appendix_ipv6}

In the IPv6 Internet topology each node represens an Autonomous System. The data was collected and developed by the Archipelago active measurement infrastructure (ARK) and CAIDA~\cite{ark2009}. 
Autonomous Systems are constituents of the Internet infrastructure that are managed by a single entity and exchange traffic between them. 
Connections in the topology represent relationships between different Autonomous Systems to facilitate the exchange of traffic. 
The dataset~\cite{as_topo_data_ipv6} consists of Autonomous Systems that route information using the IPv6 protocol. The topology used in~\cite{geometry:multilayer} was obtained by joining the connections during the first $15$ days of January $2015$ taken from~\cite{as_topo_data_ipv6}. The network has $N=5163$ nodes, a power law degree distribution with exponent $\gamma=2.1$, and average degree of $\left< k \right> = 5.21$, and an average mean local clustering coefficient of $c = 0.55$. The processed network data 
and its hyperbolic mapping were taken from publication~\cite{geometry:multilayer}. The hyperbolic mapping was obtained using the \emph{HyperMap} method~\cite{frag:hypermap,frag:hypermap_cn}. An implementation of this method is publicly available at~\cite{hypermap_code}.
\label{sec_appendix_hypermap}

\section{Localized initial conditions in the IPv6 topology}
\label{app_sec_inet_initial}

To assign localized initial conditions in the real Internet, due to the non-uniform distribution of angular coordinates, we cannot proceed like in the case of synthetic networks described by Eq.~\eqref{eqn_localized}. This procedure would lead to significantly different numbers of initial cooperators depending on which section of the angular space one selects. Instead, we perform as follows. First, we order all nodes according to their angular coordinate. Then, we randomly select a node, $i$, and simply assign the nodes in the interval $[i, i + N C_0]$ to cooperators. The ordering then ensures that they all lie within an angular sector. We choose node $i$ randomly so that the set of initial cooperators differs between different realizations.

\end{document}